\documentclass[pla,superscriptaddress,showpacs,twocolumn,10pt]{revtex4}
\usepackage{amssymb}
\usepackage{amsmath}
\usepackage{graphicx}

\begin{document}%

\newcommand{\ket}[1]{|#1\rangle}
\newcommand{\bra}[1]{\langle#1|}
\newcommand{\lr}\longrightarrow
\newcommand{\ra}\rightarrow
\newcommand{\tr}{{\rm Tr}}
\newcommand{\fsp}{{\rm span}}
\newcommand{\fsup}{{\rm supp}}
\newcommand{\fdg}{{\rm diag}}
\newcommand{\Mix}{{\rm Mix}}

\newtheorem{thm}{Theorem}
\newtheorem{Prop}{Proposition}
\newtheorem{Coro}{Corollary}
\newtheorem{Lemma}{Lemma}
\newtheorem{Def}{Definition}

\title{Unambiguous discrimination of mixed quantum states}
\author{Chi Zhang}
\email{zangcy00@mails.tsinghua.edu.cn}
\author{Yuan Feng}
\email{fengy99g@mails.tsinghua.edu.cn}
\author{Mingsheng Ying}
\email{yingmsh@tsinghua.edu.cn}\affiliation{State Key Laboratory
of Intelligent Technology and Systems, Department of Computer
Science and Technology Tsinghua University, Beijing, China,
100084}
\date{\today}\begin{abstract}
The problem of unambiguous discrimination between mixed quantum
states is addressed by isolating the part of each mixed state
which has no contribution to discrimination and by employing the
strategy of set discrimination of pure states. A necessary and
sufficient condition of unambiguous mixed state discrimination is
presented. An upper bound of the efficiency is also derived.
\end{abstract}

\maketitle

\section{Introduction}
Quantum state discrimination is an essential problem in quantum
information theory. It was first considered for pure states in the
literature. One strategy of quantum state discrimination is that
one is asked to distinguish the given states without error, but a
non-zero probability of inconclusive answer is allowed. Such a
strategy is called unambiguous discrimination. Unambiguous
discrimination for pure states was originally addressed by
Ivanovic \cite{IV}, and then Dieks \cite{DI} and Peres \cite{PE},
all of whom focused on the case in which the two states have equal
prior probabilities. Their results have been extended to the case
of two pure states with unequal prior probability by Jaeger and
Shimony \cite{JS}. Furthermore, Chefles \cite{CH} considered the
general case of $n$ quantum pure states and pointed out that they
can be unambiguously discriminated if and only if they are
linearly independent. It was observed by Sun \textit{et al}
\cite{SZ} that finding the optimal unambiguous discrimination,
which has the maximal success probability, can be reduced to a
semi-definite programming (SDP) problem. In Ref.\cite{Eldar1},
Eldar developed a set of necessary and sufficient conditions for
optimal unambiguous discrimination. Zhang \textit{et al}
\cite{ZF} and Feng \textit{et al} \cite{FZ} derived two lower
bounds on the inconclusive probability of unambiguous
discrimination among $n$ pure states.

Recently, the unambiguous discrimination among mixed states
attracted a lot of attention in the quantum information community.
Rudolph \textit{et al} \cite{RU} derived a lower and upper bound
on the maximal success probability of unambiguous discrimination
of two mixed states. Raynal \textit{et al} \cite{RA} developed a
method which eliminate unwanted subspaces of states to reduce the
discrimination of mixed states to that of some simpler states
which have the same rank. Fiurasek and Jezek \cite{FJ} introduced
some necessary and sufficient conditions on the optimal
unambiguous discrimination and presented a numerical method. Eldar
\cite{Eldar2} also obtained some necessary and sufficient
conditions for optimal unambiguous discrimination. Feng \textit{et
al} \cite{Feng} gave a necessary and sufficient condition for
unambiguous discrimination, and derived a series of lower bounds
on the inconclusive probability for unambiguous discrimination.
Another necessary and sufficient condition is mentioned in
Ref.\cite{LOCC} by Chefles.

The problem of unambiguously discriminating mixed quantum states
can be formally stated as follows. A quantum system is secretly
prepared in one of $m$ mixed states $\{ \rho_{i} \}$, $i=1,
\ldots, m$. Each $\rho_{i}$ is in the $n$-dimensional Hilbert
space and $n \geq m$. The strategy to discriminate the states is
to design a general measurement which consists of linear operators
$\Pi_{i}$, $i=0,1,\dots,m$, satisfying $\sum_{i=0}^{m}\Pi_{i}=I$.
If for any $i \neq j$, $i \neq 0$, $\tr(\Pi_{i}\rho_{j})=0$, then
when outcome $i$ $(i \neq 0)$ is observed, one may claim with
certainty that the system is originally in the state $\rho_{i}$,
and occurrence of outcome 0 means that the identification fails
to give a report.

As a natural generalization of discrimination of pure states,
Zhang \textit{et al} \cite{Set} considered the problem of set
discrimination. Suppose that a quantum system is secretly prepared
in one of some pure states which have been partitioned into a
family of disjoint sets. The aim of set discrimination is to tell
which set the state of the system is in.

It is well-known that mixed states can be regarded as ensembles of
pure states. So a natural question is whether we can perform mixed
state discrimination by using set discrimination of pure states.
The main purpose of the current paper is to give a positive answer
to this question. First, we give a method to divide each
$\rho_{i}$ into two parts. The first parts are intrinsically
undifferentiated, in the sense that they cannot unambiguously
discriminated, while the second parts of these states can be
further separated into linearly independent pure states. This
allows us to reduce a problem of discriminating mixed states into
a corresponding problem of set discrimination for pure states.
Then we are able to give a necessary and sufficient condition on
unambiguous discrimination of mixed states by using some related
results in Ref.\cite{Set}. Furthermore, the efficiency of mixed
state discrimination is carefully examined, and we present an
upper bound on it. Finally, the concept of set discrimination is
generalized to the case of mixed states. It is concluded that for
mixed states, the problem of unambiguous set discrimination is
just equivalent to that of state discrimination. This is quite
different from the case of pure states.

\section{Division of mixed states}\label{MI}
As we know, a mixed state can be given rise from some ensembles of
pure states. If a measurement can unambiguously discriminate the
ensembles for the mixed states we consider, it can also
discriminate the corresponding mixed states. As a result, at the
first glance, it seems that the technique of set discrimination
of pure states can be directly used to deal with discrimination
of mixed states. Unfortunately, it is not the case. Indeed, the
condition for a family of pure state sets to be unambiguously
discriminated, is that the span space of each set is linearly
independent to that of the others \cite{Set}. This condition is
very strict and cannot be satisfied when two mixed states is
considered to have joint support space. But lots of such states
can be unambiguously discriminated virtually.

However, things are not so dismaying. In this section, we first
divide each mixed state into two parts, such that the the first
part of each state has the support space contained in the sum of
the other states' support spaces, while the second part of each
state has the support space which intersects the sum of all other
states' support spaces only at the zero space. Then we can
concentrate our attention to the second parts of these mixed
states and the technique of set discrimination can be applied.

To prove the possibility of the division, we start with the
following lemma.

\begin{Lemma}\label{DMS}
Let $\rho$ be a density matrix with support space S. For any
subspace $M \subseteq S$, there exists a unique pair of positive
matrices, $\rho_{1}$ and $\rho_{2}$, such that

(1) $\rho_{1}+\rho_{2} = \rho$;

(2) $\fsup(\rho_{1}) = M$;

(3) $\fsup(\rho_{1}) \bigcap \fsup(\rho_{2}) = \{0\}$.
\end{Lemma}

{\it Proof.} Suppose \{$\ket{\psi_i}$, $i=1, \ldots, m$\} are
linearly independent unnormalized vectors and
$\rho=\sum_{i=1}^{m}\ket{\psi_i}\bra{\psi_i}$, which implies that
the dimension of space S is $m$. Define matrix $\Psi =
\begin{pmatrix} \ket{\psi_1} &\ldots &\ket{\psi_m}
\end{pmatrix}$.\

First we claim that there exists unitary matrix $U =
\begin{pmatrix} U_1 &U_2 \end{pmatrix}$ such that
$\fsp(\Psi U_1) = M$.

Note that $M \subseteq S = \fsp(\ket{\psi_i})$. Let $n$ be the
dimension of subspace M. There exist linearly independent vectors
$\{\ket{\varphi_1},\ldots,\ket{\varphi_n}\}$ in $C^m$ satisfying
\begin{equation}
\fsp(\Psi\ket{\varphi_i}:i=1,\ldots,n) = M.
\end{equation}

From the vectors $\{\ket{\varphi_i}\}$, we can produce by
Cram-Schmidt procedure a set $\{\ket{u_i}:i=1,\ldots,n\}$ of
normalized vectors whose span space is just the span space of
$\{\ket{\varphi_i}\}$. It is then easy to see that
$\fsp(\Psi\ket{u_i}:i=1,\ldots,n)$ is the span space of
$\{\Psi\ket{\varphi_i}:i=1,\ldots,n\}$, i.e. ,
\begin{equation}\label{l1:1}
\fsp(\Psi\ket{u_i}:i=1,\ldots,n) = M.
\end{equation}

For the normalized vectors $\{\ket{u_i}:i=1,\ldots,n\}$, there
exist normalized vectors $\{\ket{u_j}:j=n+1,\ldots,m\}$ whose span
space is orthogonal to $\fsp(\ket{u_i}:i=1,\ldots,n)$. So the
matrix $U=\begin{pmatrix}\ket{u_1} &\ldots
&\ket{u_m}\end{pmatrix}$ is a unitary matrix. Let $U_1 =
\begin{pmatrix}\ket{u_1} &\ldots &\ket{u_n}\end{pmatrix}$, and
$U_2 = \begin{pmatrix}\ket{u_{n+1}} &\ldots
&\ket{u_m}\end{pmatrix}$. Then we obtain that $\fsp(\Psi U_1) =
M$.

Next, we consider $\fsp(\Psi U_1)$ and $ \fsp(\Psi U_2)$. Because
the columns of $\Psi$, $\{\ket{\psi_i}\}$, are linearly
independent, we have
\begin{equation}\label{l1:2}
\fsp(\Psi U_1) \cap \fsp(\Psi U_2) = \{0\}
\end{equation}
from the fact that  $\fsp(U_1)\cap \fsp(U_2) = \{0\}$. Let $\rho_1
= \Psi U_1 U_1^\dagger\Psi^\dagger$, and $\rho_2 = \Psi U_2
U_2^\dagger\Psi^\dagger$. It can be easily observed that
$\fsup(\rho_1)=\fsp(\Psi U_1)$ and $\fsup(\rho_2)=\fsp(\Psi U_2)$.
From Eqs.(\ref{l1:1}) and (\ref{l1:2}), we know that $\rho_1$,
$\rho_2$ are the matrices we want.

Now, we will show that the matrices $\rho_1$ and $\rho_2$ are
unique. Assume that there are two different pairs $\{\rho_1,
\rho_2\}$, $\{\rho'_1, \rho'_2\}$ satisfying the conditions (1),
(2), (3) in this lemma, and
\begin{equation}
\rho_1 = \sum_{i=1}^{n}\ket{\psi_i}\bra{\psi_i},\ \ \rho_2 =
\sum_{i=n+1}^{m}\ket{\psi_i}\bra{\psi_i};
\end{equation}
\begin{equation}
\rho'_1 = \sum_{i=1}^{n}\ket{\psi'_i}\bra{\psi'_i},\ \ \rho'_2 =
\sum_{i=n+1}^{m}\ket{\psi'_i}\bra{\psi'_i}.
\end{equation}

Because of the unitary freedom in the ensembles for density
matrices, there must exist a unitary matrix $V$ satisfying
\begin{equation}
\ket{\psi'_i} = \sum_{j=1}^{m}v_{ij}\ket{\psi_j} =
\sum_{j=1}^{n}v_{ij}\ket{\psi_j} +
\sum_{j=n+1}^{m}v_{ij}\ket{\psi_j},
\end{equation}
where $v_{ij}$ is the ($i$, $j$) element of $V$. Since
$\fsp(\ket{\psi'_i}:i=1,\ldots,n) = M$ and
$\fsp(\ket{\psi_i}:i=n+1,\ldots,m)\cap M ={0} $, we know that
$v_{ij} = 0$ for any $1\leq i \leq n, n+1\leq j \leq m$ or $1\leq
j \leq n, n+1\leq i \leq m$, i.e.,
\begin{equation}
V = \begin{pmatrix} V_1 &0 \\ 0 &V_2\end{pmatrix}
\end{equation}
for some unitary matrices $V_1$ and $V_2$. Now, from the fact:
\begin{equation}
\begin{pmatrix}\ket{\psi_1} &\ldots &\ket{\psi_n} \end{pmatrix} =
\begin{pmatrix}\ket{\psi'_1} &\ldots &\ket{\psi'_n}
\end{pmatrix}V_1
\end{equation}
and
\begin{equation}
\begin{pmatrix}\ket{\psi_{n+1}} &\ldots &\ket{\psi_m} \end{pmatrix} =
\begin{pmatrix}\ket{\psi'_{n+1}} &\ldots &\ket{\psi'_m}
\end{pmatrix}V_2,
\end{equation}
it follows that
\begin{equation}
\begin{split}
\rho_1 & = \sum_{i=1}^{n}\ket{\psi_i}\bra{\psi_i}\\
& = \begin{pmatrix} \ket{\psi_1} &\ldots \ket{\psi_n}
\end{pmatrix}
\begin{pmatrix}\bra{\psi_1} \\ \vdots \\ \bra{\psi_n}
\end{pmatrix}\\ & = \begin{pmatrix}\ket{\psi'_1} &\ldots \ket{\psi'_n} \end{pmatrix}
V_1 V_1^+\begin{pmatrix}\bra{\psi'_1} \\ \vdots \\
\bra{\psi'_n} \end{pmatrix}\\ & = \rho'_1.
\end{split}
\end{equation}
And $\rho_2 = \rho - \rho_1 = \rho - \rho'_1 = \rho'_2$. This
completes the proof. \hfill $\Box$

We now need to introduce several notations which allow us to use
Lemma \ref{DMS} properly in the remainder of this paper.

Since the intersection of the kernels of all $\rho_i$,
$i=1,\ldots,m$, is not useful for the purpose of unambiguous
discrimination \cite{Feng}, we can assume without loss of
generality that each measurement operator $\Pi_i$, $i=1,\ldots,m$,
is in the sum of support spaces of all $\rho_i$ \cite{dSum}.

Define the space $\Mix(\rho_i)$ for each state $\rho_i$ as
follows:
\begin{equation}\label{note1}
\Mix(\rho_{i}) = \fsup(\rho_{i})\bigcap\sum^{m}_{j=1,j \neq
i}\fsup(\rho_{j}).
\end{equation}

It is obvious that for any $\rho_i$, $\Mix(\rho_i) \subseteq
\fsup(\rho_i)$. From Lemma \ref{DMS}, each $\rho_i$ can be
divided into two parts, $\tilde{\rho}_i$ and $\hat{\rho}_i$, such
that
\begin{equation}\label{note2}
\begin{split}
&\hat{\rho}_i+\tilde{\rho}_i = \rho_i;\\
&\fsup(\hat{\rho}_i) = \Mix(\rho_i);\\
&\fsup(\tilde{\rho}_i) \cap \Mix(\rho_i) = \{0\}.
\end{split}
\end{equation}

Furthermore, we define
\begin{equation}\label{2:2}
\tilde{\rho}_0 = \sum_{i=1}^m\hat{\rho}_i.
\end{equation}
Then $\tilde{\rho}_i$ is called the core of mixed state $\rho_i$,
for each $i=1,\ldots,m$.

From these definitions, it is easy to see
\begin{equation}\label{2:0}
\sum_{i=1}^m \fsup(\rho_i) = \bigoplus_{i=0}^m
\fsup(\tilde{\rho}_{i}).
\end{equation}

Moreover, we have the following lemma which establishes a close
link from discrimination measurement of $\{\rho_1,\ldots,\rho_m\}$
to $\{\tilde{\rho}_0,\ldots,\tilde{\rho}_m\}$.

\begin{Lemma}\label{EQu}
A measurement which consists of operators
$\{\Pi_{0},\ldots,\Pi_{m}\}$ can unambiguously discriminate
$\{\rho_1,\ldots,\rho_m\}$ if and only if for any $i \neq j,
i=1,\ldots,m, j=0,\ldots,m$,
\begin{equation}
\Pi_i\tilde{\rho}_j = 0,\ \ \tilde{\rho}_j\Pi_i = 0.
\end{equation}
\end{Lemma}

{\it Proof.} From Ref.\cite{Feng}, we know that any measurement
$\{\Pi_{0},\ldots,\Pi_{m}\}$ can unambiguously discriminate mixed
states $\{\rho_i:i=1,\ldots,m\}$ if and only if for any $i \neq
j$, it holds that $\Pi_i\rho_j = 0$ and $\rho_j\Pi_i = 0$.

And from the definitions given above, for any $j=1,\ldots,m$, we
have $\rho_j = \tilde{\rho}_j + \hat{\rho}_j$. Because of the
positivity of $\tilde{\rho}_j$ and $\hat{\rho}_j$, the statement
that operators $\{\Pi_i\}$ can unambiguously discriminate states
$\{\rho_j\}$ is equivalent to that for any $i \neq j$,
$i,j=1,\ldots,m$,
\begin{equation}
\Pi_i\tilde{\rho}_j = 0,\ \ \tilde{\rho}_j\Pi_i = 0,
\end{equation}
and
\begin{equation}\label{2:1}
\Pi_i\hat{\rho}_j = 0, \ \ \hat{\rho}_j\Pi_i = 0.
\end{equation}
Because of Eq.(\ref{note1}) and Eq.(\ref{note2}), for any
$i=1,\ldots,m$, it holds that
\begin{equation}
\fsup(\hat{\rho}_i) \subseteq \sum_{j \neq i}\fsup(\rho_{j}).
\end{equation}
As a result, when $i = j$, Eq.(\ref{2:1}) is also satisfied.
Further more, by the definition of $\tilde{\rho}_0$,
Eq.(\ref{2:1}) is equivalent to that $\Pi_i\tilde{\rho}_0 = 0$ and
$\tilde{\rho}_0\Pi_i = 0$, for any $i$. This completes the proof.
$\hfill$ $\Box$

With the help of above lemmas, now we are able to transform the
problem of unambiguously discriminating mixed states into that of
unambiguously discriminating sets of pure states. Such a
transformation is explicitly presented in the following theorem.

\begin{thm}\label{Center}
Let $\{\rho_1, \ldots, \rho_m\}$ be a set of mixed states, and let
$\{\tilde{\rho}_0, \tilde{\rho}_1, \ldots, \tilde{\rho}_m\}$ be
their ''core''s, which has been defined by Eqs.(\ref{note2}) and
(\ref{2:2}). Furthermore, suppose that for each $0\leq i\leq n$,
$S_i$ is a set of pure states, such that
\begin{equation}
\tilde{\rho}_i = \sum_{\ket{\psi}\in
S_i}p_{\ket{\psi}}\ket{\psi}\bra{\psi},
\end{equation}
and $p_{\ket{\psi}} > 0$ for each $\ket{\psi}\in S_i$. Then the
measurement $\{\Pi_{\ast},\Pi_{1},\ldots,\Pi_{m}\}$ unambiguously
discriminate $\{\rho_1, \ldots, \rho_m\}$ if and only if for some
operator $\Pi_0$, the measurement
$\{\Pi_{\ast}-\Pi_{0},\Pi_{0},\Pi_{1},\ldots,\Pi_{m}\}$
unambiguously discriminate sets $\{S_0, S_1, \ldots, S_m\}$, where
both $\Pi_{\ast}$ and $\Pi_{\ast}-\Pi_{0}$ are introduced for the
inconclusive reports.
\end{thm}

{\it Proof.} ``$\Longrightarrow$". From the conditions, for any
$i=0,\ldots,m$, $S_i = \{ \left| \psi_{ik} \right\rangle\}$ gives
rise to the density matrix $\tilde{\rho}_{i}$ with probabilities
$\{q_{ik}\}$, which means
\begin{equation}
\tilde{\rho}_i = \sum_{k}q_{ik}\ket{\psi_{ik}}\bra{\psi_{ik}},
\end{equation}
where $\sum q_{ik}$ is not necessarily $1$.

Let $\{\Pi_{\ast}, \ldots, \Pi_{m}\}$ be any measurement which can
unambiguously discriminate the mixed states, i.e.,
$\tr(\Pi_j\rho_i) = 0$, for any $i \neq j$.

From Eq.(\ref{2:1}), we have, for any $i,j = 1,\ldots, m$,
\begin{equation}
\tr(\Pi_{j}\hat{\rho}_{i}) = 0.
\end{equation}
So,
\begin{equation}\label{a}
\begin{split}
\tr(\Pi_{j}\rho_{i}) & = \tr(\Pi_{j}\tilde{\rho}_{i})\\ & =
\sum_{k=1}^{n_{i}} q_{ik}\left\langle
\psi_{ik}\right|\Pi_{j}\left| \psi_{ik} \right\rangle
\end{split}
\end{equation}
Because every element in Eq.(\ref{a}) is nonnegative, when $i \neq
j$, $\tr(\Pi_j\rho_i) = 0$, it holds that $\left\langle
\psi_{ik}\right|\Pi_{j}\left| \psi_{ik} \right\rangle =0$, for
any $k$.

Furthermore, since $\Pi_j\tilde{\rho}_0 = 0$ and
$\tilde{\rho}_0\Pi_j = 0$, $\left\langle
\psi_{0k}\right|\Pi_{j}\left| \psi_{0k} \right\rangle =0$, for
any $k,j$. It is easy to see that the measurement $\{\Pi_{\ast},
0, \Pi_{1},\ldots, \Pi_{m}\}$ can unambiguously discriminate the
sets, leaving the efficiency of identifying set $S_0$ zero.

``$\Longleftarrow$". Let $\{\Pi_{\dagger}, \Pi_{0}, \ldots,
\Pi_{m}\}$ be any measurement operators which can unambiguously
discriminate the sets. Because $\fsup(\hat{\rho}_i) \subseteq
\fsup(\tilde{\rho}_0) = \fsp(S_0)$, we have
\begin{equation}
\tr(\Pi_{j}\hat{\rho}_i) = 0,
\end{equation}
for any $i,j = 1, \ldots, m$. So
\begin{equation}\label{2:10}
\begin{split}
\tr(\Pi_{j}\rho_{i}) & = \tr(\Pi_{j}\tilde{\rho}_{i})\\ & =
\sum_{k=1}^{n_{i}} q_{ik}\left\langle
\psi_{ik}\right|\Pi_{j}\left| \psi_{ik} \right\rangle\\ & =
\left(\sum_{k=1}^{n_{i}} q_{ik}\gamma_{ik}\right) \delta_{ij},
\end{split}
\end{equation}
where $\gamma_{ik}$ is the efficiency of identifying $\left|
\psi_{ik} \right\rangle$ in the set unambiguous discrimination.
In this way, the measurement
$\{\Pi_{\ast}=\Pi_{\dagger}+\Pi_{0},\Pi_{1},\ldots,\Pi_{m}\}$ can
unambiguously discriminate the mixed states
$\{\rho_1,\ldots,\rho_m\}$. \hfill $\Box$

Then a corollary, which presents a necessary and sufficient
condition of unambiguous discrimination between mixed states, can
be derived.

\begin{Coro}\label{NSC}
The states $\{\rho_{i}:i=1,\ldots,m\}$ can be unambiguously
discriminated if and only if for any $i=1,\ldots,m$, the core of
$\rho_i$, $\tilde{\rho}_i$, is not zero.
\end{Coro}

{\it Proof.} ``$\Longrightarrow$". Suppose that $\tilde{\rho}_{i}
= 0$, for some $1 \leq i \leq m$. From Eq.(\ref{2:1}), we know
that for any unambiguous discrimination operators \{$\Pi_i$\}
\begin{equation}
\tr(\Pi_{i}\rho_{i})= \tr(\Pi_{i}\tilde{\rho}_{i}) = 0,
\end{equation}
which contradicts the definition of unambiguous discrimination.

``$\Longleftarrow$". Let $\{S_i:i=0,\ldots,m\}$ be any state sets
satisfying that $S_i$ can give rise to state $\tilde{\rho}_i$,
$i=0,\ldots,m$. From Eq.(\ref{2:0}), $\{S_i\}$ are linearly
independent. By using the results presented in Ref.\cite{Set}, we
can design a measurement $\{\Pi_{\dagger}, \Pi_{0}, \ldots,
\Pi_{m}\}$ to unambiguously discriminate them. From Theorem
\ref{Center}, the measurement operators
$\{\Pi_{\ast}=\Pi_{\dagger}+\Pi_{0},\Pi_{1},\ldots,\Pi_{m}\}$ can
discriminate the mixed states $\{\rho_1,\ldots,\rho_m\}$
unambiguously. \hfill $\Box$

To conclude this section, we compare the results obtained in this
section with some related works. It is easy to see that when the
states are all pure, the condition given in the above corollary
degenerates to the known necessary and sufficient condition for
pure state discrimination presented in Ref.\cite{CH}. In
Ref.\cite{Feng}, Feng \textit{et al} pointed out that mixed states
$\{\rho_1, \ldots, \rho_m\}$ can be unambiguously discriminated if
and only if for any $i \neq j$, $\fsp(S) \neq \fsp(S_i)$, where $S
= \{\rho_1, \ldots, \rho_m\}$ and $S_i = S\backslash \{\rho_i\}$.
From Eq.(\ref{2:0}), it is easy to see
$\fsp(S_i)\oplus\fsup(\tilde{\rho}_i) = \fsp(S)$. So, the the
condition in Corollary 1 and that presented in Ref.\cite{Feng} are
equivalent. In Ref.\cite{LOCC}, Chefles also gave a necessary and
sufficient condition for mixed state discrimination which requires
that the orthogonal complements of $\sum_{j \neq i}\fsup(\rho_j)$
is not a subspace of the orthogonal complement of $\fsup(\rho_i)$,
for any $i$. The condition is the same as that $\fsup(\rho_i)
\nsubseteq \sum_{j \neq i} \fsup(\rho_j)$ and can be easily
reduced to the condition given in Ref.\cite{Feng}, which have been
proved equivalent to the condition in Corollary \ref{NSC}.

\section{Efficiency of unambiguous discrimination}\label{EF}

In the last section, we have established a close link between
mixed state discrimination and set discrimination of pure states.
In this section, we are going to evaluate efficiency of
unambiguous discrimination of mixed states by using this link.

In Ref. \cite{Set}, Zhang \textit{et al} showed that the sets
$\{S_0, \ldots, S_1\}$ can be unambiguously discriminated with
efficiency $\{\gamma_{ik}:i=0,\ldots,m;k=1,\ldots,n_i\}$ if and
only if there are matrices $\{\Gamma_0,\ldots,\Gamma_m\}$ such
that $\Gamma_i$ is a $n_i\times n_i$ positive matrix with the
$k$th diagonal entry being $\gamma_{ik}$, and the matrix
\begin{equation}\label{3:1}
X-\Gamma\geq 0,
\end{equation}
where $X=\begin{pmatrix}\left\langle \psi_{ik}\mid \psi_{jl}
\right \rangle\end{pmatrix}$, and $\Gamma =
diag(\Gamma_0,\ldots,\Gamma_m)$. Thus, the problem of evaluating
the optimal efficiency of set discrimination of pure states is
connected to a problem of semi-positive programming.

On the other hand, it is shown in Theorem \ref{Center} that the
measurement operators $\{\Pi_{*}, \Pi_{1}, \ldots,\Pi_m\}$ can
unambiguously discriminate states $\{\rho_1,\ldots,\rho_m\}$, if
and only if there exists $\Pi_0$ to make measurement operators
$\{\Pi_{*}-\Pi_0,\Pi_0,\ldots,\Pi_m\}$ unambiguously discriminate
sets $\{S_0,\ldots,S_m\}$ which give rise to the density operators
$\{\tilde{\rho}_0,\ldots,\tilde{\rho}_m\}$ defined by Eqs.
(\ref{note1}) and (\ref{2:2}). Combining these results enables us
to derive an estimation of the efficiency of unambiguous
discrimination of mixed states

Let
\begin{equation}\label{3:3}
\begin{split}
\tilde{\rho}_i &=
\sum_{k=1}^{n_i}q_{ik}\ket{\psi_{ik}}\bra{\psi_{ik}}\\ &=
\sum_{k=1}^{n_i}\ket{\tilde{\psi}_{ik}}\bra{\tilde{\psi}_{ik}},
\end{split}
\end{equation}
where $i=0,\ldots,m$, $\{\ket{\psi_{ik}}\}$ are pure states in set
$S_i$, and
$\ket{\tilde{\psi}_{ik}}=\sqrt{q_{ik}}\ket{\psi_{ik}}$. Let
$\gamma_i$ is the the success probability of identifying $\rho_i$
in the mixed state discrimination. From Eq.(\ref{2:10}), we have
\begin{equation}
\gamma_i = \left(\sum_{k=1}^{n_{i}} \gamma_{ik}q_{ik}\right),
\end{equation}
where $\{q_{ik}\}$ are defined in Eq.(\ref{3:3}), and
$\gamma_{ik}$ is the efficiency of discriminating
$\ket{\psi_{ik}}$ in set $S_i$, which can be determined by
Eq.(\ref{3:1}). So, we have
\begin{equation}
\gamma_i = \tr\left(Q_i\Gamma_i{Q_i}\right),
\end{equation}
where $Q_i = \fdg(\sqrt{q_{ik}})$, $k=1,\ldots,n_i$, and
$\Gamma_i$ is a $n_i\times n_i$ positive matrix with the $k$th
diagonal entry being $\gamma_{ik}$. Let $Q =
\fdg(Q_0,\ldots,Q_m)$, $\tilde{X} = QXQ$, $\tilde{\Gamma} =
Q\Gamma{Q}$. Since $Q$ is a positive diagonal matrix, from
Eq.(\ref{3:1}), we obtain
\begin{equation}\label{rel}
\tilde{X}-\tilde{\Gamma} \geq 0.
\end{equation}
It is easy to see that $\tilde{X} =
\begin{pmatrix}\left\langle
\tilde{\psi_{ik}}\mid \tilde{\psi_{jl}} \right
\rangle\end{pmatrix}$, $\tilde{\Gamma} =
\fdg(\tilde{\Gamma_0},\ldots,\tilde{\Gamma_m})$, and
$\tilde{\Gamma_i}$ is a $n_i\times{n_i}$ positive matrix with its
trace, $\tr(\tilde{\Gamma_i})$, as the efficiency of
discriminating $\rho_i$. Moreover, if measurement operators
$\{\Pi_{\ast}-\Pi_{0},\Pi_{0},\ldots,\Pi_m\}$ can unambiguously
discriminate $\{S_0,\ldots,S_m\}$, so do
$\{\Pi_{\ast},0,\Pi_1,\ldots,\Pi_m\}$, leaving the success
probability of $S_0$ zero. Thus, let $\tilde{\Gamma}_0 = 0$, the
Eq.(\ref{rel}) is also satisfied.

From the above argument, we have the following theorem.
\begin{thm}\label{Eff}
The mixed states $\{\rho_{i}:i=1,\ldots,m\}$, whose core,
$\{\tilde{\rho}_i:i=0,\ldots,m\}$, can be expressed as
Eq.(\ref{3:3}), would be unambiguously discriminated with
efficiency $\{\gamma_{i}:i=1,\ldots,m\}$ if and only if
\begin{equation}
\tilde{X}-\tilde{\Gamma} \geq 0,
\end{equation}
and
\begin{equation}
\tilde{\Gamma} \geq 0,
\end{equation}
where $\tilde{X} =
\begin{pmatrix}\left\langle
\tilde{\psi_{ik}}\mid \tilde{\psi_{jl}} \right
\rangle\end{pmatrix}$, $\tilde{\Gamma} =
\fdg(0,\tilde{\Gamma_1},\ldots,\tilde{\Gamma_m})$, and $\gamma_{i}
= \tr(\tilde{\Gamma_{i}})$.
\end{thm}

Notice that there are many different ensembles which can give rise
to $\tilde{\rho}_i$. This implies that the matrix $\tilde{X}$ in
the above theorem is not unique. We can prove that, however, the
discrimination efficiency derived from different $\tilde{X}$ are
the same.

For different ensembles $\{\ket{\tilde{\psi_{ik}}}\}$ and
$\{\ket{\tilde{\varphi_{il}}}\}$ which give rise to the same
density operator $\tilde{\rho}_i$,
\begin{equation}
\begin{split}
\tilde{\rho}_i &=
\sum_{k}\ket{\tilde{\psi_{ik}}}\bra{\tilde{\psi_{ik}}}\\
&=\sum_{l}\ket{\tilde{\varphi_{il}}}\bra{\tilde{\varphi_{il}}}
\end{split}
\end{equation}
Because of the unitary freedom for ensembles, we have
\begin{equation}
\ket{\tilde{\psi_{ik}}} =
\sum_{l}u_{kl}\ket{\tilde{\varphi_{il}}},
\end{equation}
where $u_{kl}$ is the $(k,l)$ entry of a unitary matrix $U_i$. Let
$\tilde{X} =
\begin{pmatrix}\left\langle
\tilde{\psi}_{ik_1}\mid \tilde{\psi}_{jk_2} \right
\rangle\end{pmatrix}$, and $\tilde{Y} =
\begin{pmatrix}\left\langle
\tilde{\varphi}_{il_1}\mid \tilde{\varphi}_{jl_2} \right
\rangle\end{pmatrix}$, we have
\begin{equation}
\tilde{X} = U^\dagger\tilde{Y}U,
\end{equation}
where $U = \fdg(I,U_1,\ldots,U_m)$ which is also a unitary matrix.

For any $\tilde{\Gamma}$ , let $\tilde{\Lambda} =
U^\dagger\tilde{\Gamma}U$. Then $\tilde{\Gamma} \geq 0$ if and
only if $\tilde{\Lambda} \geq 0$, and $\tilde{X} - \tilde{\Gamma}
\geq 0$ if and only if $\tilde{Y} -\tilde{\Lambda} \geq 0$. The
efficiency of discriminating state $\rho_i$ in the second way is
\begin{equation}
\tr(\Lambda_i) = \tr(U_i^\dagger\Gamma_i{U_i}),
\end{equation}
which is the same as that in the first way, $\tr(\Gamma_i)$. As a
consequence, different ensembles for the same density operators do
not introduce any difference in the efficiency of unambiguous
discrimination.

It should be pointed out that, in Ref.\cite{YMB}, a result
considering optimal efficiency of mixed state discrimination, has
been derived by Eldar \textit{et al}. However, the method she used
is derived directly from the conditions of the measurement
operators, which is quite different from ours. In this paper, we
first combine the concept of set discrimination and the concept of
mixed state discrimination.

\section{Upper bound of unambiguous discrimination efficiency}
For pure states, an upper bound for the maximal success
probability of unambiguous discrimination was found in Refs.
\cite{ZF} and \cite{FZ}. In this section, we derive an upper bound
on the success probability of unambiguous discrimination among $m$
mixed states $\{\rho_i\}$, with prior probability $\{\eta_i\}$.

First,we need to prove a lemma for positive matrix.
\begin{Lemma}\label{PM}
For any positive matrix $\begin{pmatrix}A& B \\B^\dagger & C
\end{pmatrix}$, it holds that
\begin{equation}
\tr(A)\tr(C) \geq \tr^2(\sqrt{B^\dagger B}).
\end{equation}
\end{Lemma}

{\it Proof.} Because $\begin{pmatrix} A  &B\\
B^\dagger &C\end{pmatrix}$ is positive, there must be matrices
$Q_1$ and $Q_2$ satisfying that
\begin{equation}
\begin{pmatrix} A  &B\\ B^\dagger &C\end{pmatrix} =
\begin{pmatrix}Q_1^\dagger \\ Q_2^\dagger\end{pmatrix}
\begin{pmatrix}Q_1 &Q_2\end{pmatrix},
\end{equation}
i.e. $A = Q_1^\dagger Q_1$, $B = Q_1^\dagger Q_2$, and $C =
Q_2^\dagger Q_2$.

Furthermore, it holds that
\begin{equation}
\begin{split}
\tr(A) \tr(C) & = \tr(Q_1^\dagger Q_1) \tr(Q_2^\dagger Q_2)\\ & =
\tr(Q_1Q_1^\dagger) \tr(Q_2Q_2^\dagger)\\ & =
\tr(\begin{pmatrix}Q_1&0\end{pmatrix} UU^\dagger \left(\begin{matrix}Q_1^\dagger \\
0\end{matrix}\right)) \tr(Q_2Q_2^\dagger),
\end{split}
\end{equation}
for any unitary matrix $U$.

Using Cauchy inequality, we have
\begin{equation}
\begin{split}
\tr(A) \tr(C) &
\geq \begin{vmatrix} \tr(\begin{pmatrix}Q_1&0\end{pmatrix}U  \left(\begin{matrix}Q_2^\dagger \\
0\end{matrix}\right)) \end{vmatrix}^2 \\
& = \begin{vmatrix} \tr(\begin{pmatrix}Q_1 &0\end{pmatrix}\begin{pmatrix}Q_2^\dagger \\
0\end{pmatrix}U \end{vmatrix}^2 \\
& = \begin{vmatrix}\tr(\begin{pmatrix}Q_2^\dagger Q_1&0
\\ 0 &0\end{pmatrix}U) \end{vmatrix}^2 \\
& = \begin{vmatrix} \tr(\begin{pmatrix}B^\dagger &0 \\ 0
&0\end{pmatrix}U)
\end{vmatrix}^2.
\end{split}
\end{equation}

Notice that for any operator M,
\begin{equation}
\tr\begin{vmatrix}M\end{vmatrix} = max_{V}
\begin{vmatrix} \tr(M V) \end{vmatrix},
\end{equation}
where the maximum is taken over all unitary matrices $V$. It
follows that
\begin{equation}
\begin{split}
\tr(A) Tr(C) & \geq \tr^2(\begin{vmatrix}\begin{pmatrix}B^\dagger &0 \\
0 &0\end{pmatrix}\end{vmatrix}) \\
& = \tr^2(\sqrt{\left(\begin{matrix}B^\dagger B &0 \\
0 &0\end{matrix}\right)})\\ & = \tr^2(\sqrt{B^\dagger B}).
\end{split}
\end{equation}
So the proof is completed. \hfill $\Box$

Using this lemma, we are able to give an upper bound for the
efficiency of mixed state unambiguous discrimination.
\begin{thm}
For any unambiguous discrimination of mixed states
$\{\rho_1,\ldots,\rho_m\}$ with prior probabilities
$\{\eta_1,\ldots,\eta_m\}$, the efficiency $P$ satisfies
\begin{equation}
P \leq \sum_{i=1}^{m} \eta_{i}\tr(\tilde{\rho}_i) -
\sqrt{\dfrac{m}{m-1}\sum_{i \neq
j}\eta_{i}\eta_{j}F^{2}(\tilde{\rho}_i,\tilde{\rho}_j)},
\end{equation}
where $\tilde{\rho}_i$ is the "core'' of state $\rho_i$.
\end{thm}

{\it Proof.} According to section \ref{EF}, we suppose that
$q_{ik} = \left\langle \tilde{\psi_{ik}}\mid \tilde{\psi_{ik}}
\right \rangle$, where $\ket{\tilde{\psi_{ik}}}$ is defined in
Eq.(\ref{3:3}). From the analysis in section \ref{EF}, we know
that inconclusive probability satisfies
\begin{equation}
\begin{split}
P_0 & = 1 - \sum_{i=1}^m \eta_{i}\tr(\Gamma_i)\\ & = \sum_{i=1}^m
\eta_{i}(1-\tr(\Gamma_i))\\ & = \sum_{i=1}^m
\eta_{i}\sum_{k=1}^{n_i}( q_{ik} - \gamma_{ik} ) + \sum_{i=1}^m
\eta_{i}( 1- \sum_{k=1}^{n_i} q_{ik}),
\end{split}
\end{equation}
where $\gamma_ik$ is the $k$th diagonal entry of matrix
$\Gamma_i$.

Let $\Psi_i=\begin{pmatrix}\ket{\tilde{\psi_{i1}}} &\ldots
&\ket{\tilde{\psi_{in_i}}} \end{pmatrix}$. So the $(i,j)$ block in
matrix $\tilde{X}$ is $\Psi_i^\dagger\Psi_j$. Because
$\tilde{X}-\tilde{\Gamma} \geq 0$, from Lemma \ref{PM}, we have
\begin{equation}
\sum_{k=1}^{n_i}(q_{ik} - \gamma_{ik} ) \sum_{l=1}^{n_j}(q_{ik} -
\gamma_{jl}) \geq
\tr^2(\sqrt{\Psi_j^\dagger\Psi_i\Psi_i^\dagger\Psi_j}).
\end{equation}
Notice that $\tilde{\rho}_i = \Psi_i\Psi_i^\dagger $ and
$\tilde{\rho}_j = \Psi_j\Psi_j^\dagger$. So, we have
\begin{equation}
\sqrt{\rho\tilde{}_j} = \begin{pmatrix}\Psi_j &0\end{pmatrix}
 U = U^\dagger \begin{pmatrix}\Psi_j^\dagger
\\ 0 \end{pmatrix},
\end{equation}
\begin{equation}
\sqrt{\tilde{\rho}_j} \tilde{\rho}_i \sqrt{\tilde{\rho}_j} =
U^\dagger
\begin{pmatrix}\Psi_j^\dagger\tilde{\rho}_i\Psi_j &0 \\ 0 &0
\end{pmatrix} U.
\end{equation}

In this way, we know that the nonzero eigenvalues of
$\sqrt{\tilde{\rho}_j} \tilde{\rho}_i \sqrt{\tilde{\rho}_j}$ are
the same as the nonzero eigenvalues of
$\Psi_j^\dagger\tilde{\rho}_i\Psi_j$. That means
\begin{equation}
\tr(\sqrt{\Psi_j^\dagger\Psi_i\Psi_i^\dagger\Psi_j}) =
\tr(\sqrt{\tilde{\rho}_j} \tilde{\rho}_i \sqrt{\tilde{\rho}_j}) =
F(\tilde{\rho}_i,\tilde{\rho}_j),
\end{equation}
where $F$ stands for fidelity. So it follows that
\begin{equation}
\sum_{k=1}^{n_i}(q_{ik} - \gamma_{ik} ) \sum_{l=1}^{n_j}(q_{ik} -
\gamma_{jl}) \geq F^2 (\tilde{\rho}_i,\tilde{\rho}_j).
\end{equation}
Let $x_i = \sum_{k=1}^{n_i}(q_{ik} - \gamma_{ik} )$. For any $i
\neq j$, it holds that
\begin{equation}
\begin{split}
x_i  x_j & = \eta_i\eta_j\sum_{k=1,l=1}^{n_i,n_j} (q_{ik} -
\gamma_{ik})(q_{jl} - \gamma_{jl})\\ & \geq
\eta_i\eta_jF^2(\tilde{\rho}_i,\tilde{\rho}_j).
\end{split}
\end{equation}

By using Cauchy inequality
\begin{equation}
\left(\sum_{i=1}^m x_i\right)^2 \geq \frac{m}{m-1}\sum_{i \neq
j}x_ix_j,
\end{equation}
we obtain
\begin{equation}
\begin{split}
P_0  = & \sqrt{\left(\sum_{i=1}^m x_i\right)^2} + \sum_{i=1}^m
\eta_i(1-\sum_{k=1}^{n_i} q_{ik})\\  \geq &
\sqrt{\frac{m}{m-1}\sum_{i\neq j}x_ix_j} + \sum_{i=1}^m
\eta_i(1-\sum_{k=1}^{n_i} q_{ik})\\  \geq &
\sqrt{\frac{m}{m-1}\sum_{i\neq j}\eta_i\eta_j F^2
(\tilde{\rho}_i,\tilde{\rho}_j)} \\ & +
\sum_{i=1}^m\eta_i(1-\sum_{k=1}^{n_i} \left\langle
\tilde{\psi}_{ik}\mid \tilde{\psi}_{ik} \right \rangle).
\end{split}
\end{equation}
Because of the definition of $\tilde{\rho}_i$, we know that
$\sum_{k=1}^{n_i} \left\langle \tilde{\psi}_{ik}\mid
\tilde{\psi}_{ik} \right \rangle = \tr( \tilde{\rho}_i )$. So,
\begin{equation}
P_0 \geq \sqrt{\dfrac{m}{m-1} \sum_{i \neq j}
\eta_{i}\eta_{j}F^2(\tilde{\rho}_i,\tilde{\rho}_j)} + \sum_{i=1}^m
\eta_{i}( 1 - \tr(\tilde{\rho}_i) ),
\end{equation}
and
\begin{equation}
\begin{split}
P & = 1 - P_0\\
& \leq \sum_{i=1}^{m} \eta_{i}\tr(\tilde{\rho}_i) -
\sqrt{\dfrac{m}{m-1}\sum_{i \neq
j}\eta_{i}\eta_{j}F^{2}(\tilde{\rho}_i,\tilde{\rho}_j)}.
\end{split}
\end{equation} \hfill
$\Box$

\section{Set discrimination of mixed states}
In section \ref{MI}, we point out that the unambiguous
discrimination of mixed states can be performed by unambiguous set
discrimination of pure states. Conversely, an interesting problem
is whether we can perform unambiguous set discrimination by
discriminating mixed states. For this purpose, we first generalize
the concept of set discrimination.

Suppose $S_i = \{\sigma_i^k:k=1,\ldots,n_i\}$ is a set of mixed
quantum states and $S_i \bigcap S_j = \emptyset$ for any $1\leq
i,j\leq m$ with $i \neq j$. Assume that a quantum system is
secretly prepared in one of the mixed states in $\bigcup_{i=1}^m
S_i$. We intend to tell which set of $\{S_i\}$ the unknown state
is in. Obviously, when each set $S_i$ is a singleton, this problem
reduces to mixed state discrimination. A surprising fact is that
any unambiguous set discrimination can be reduced to unambiguous
mixed state discrimination, as the following theorem states.
\begin{thm}\label{Set}
For any mixed quantum states $\sigma_i^k$, $i = 1,\ldots,m$, $k =
1,\ldots,n_i$, with prior probability $\eta_{ik}$, the problem of
unambiguously discriminating the sets $S_i =
\{\sigma_i^{k}:k=1,\ldots,n_i\}$, $i = 1,\ldots,m$, is equivalent
to that of unambiguously discriminating states $\rho_i =
\sum_{k=1}^{n_i}\frac{\eta_{ik}}{\eta'_i}\sigma_i^k$,, $i =
1,\ldots,m$, in the sense that they have the same efficiency,
where $\eta'_i = \sum_{k=1}^{n_i}\eta_{ik}$ is the prior
probability of $\rho_i$.
\end{thm}

{\it Proof.} For any measurement operators
$\{\Pi_i:i=0,\ldots,m\}$, they can unambiguously discriminate the
sets $\{S_i\}$ if and only if for any $i,j=1,\ldots,m$, $i \neq
j$,
\begin{equation}
\tr(\sigma_i^k\Pi_j) = 0.
\end{equation}

On the other hand, for any $\rho_i =
\sum_{k=1}^{n_i}\frac{\eta_{ik}}{\eta'_i}\tr(\sigma_i^k\Pi_j)$, it
holds that
\begin{equation}
\tr(\rho_i\Pi_j) =
\sum_{k=1}^{n_i}\frac{\eta_{ik}}{\eta'_i}\tr(\sigma_i^k\Pi_j).
\end{equation}

Because $\frac{\eta_{ik}}{\eta'_i} > 0$, $\sigma_i^k \geq 0$, and
$\Pi_j
> 0 $, we know that $\tr(\rho_i\Pi_j) = 0 $ if and only if for any $k$,
$\tr(\sigma_i^k\Pi_j) = 0$. So a necessary and sufficient
condition for the measurement operators $\{\Pi_i:i=1,\ldots,m\}$
to unambiguously discriminate $\{\rho_i:\rho_i =
\sum_{k=1}^{n_i}\frac{\eta_{ik}}{\eta'_i}\sigma_i^k\}$ is that for
any $i \neq j$, $\tr(\sigma_i^k\Pi_j)=0$. It is just the necessary
and sufficient condition which ensures that $\{S_i\}$ can be
unambiguously discriminated.

Furthermore, the measurement operators $\{\Pi_i\}$ can
unambiguously discriminate mixed states $\{\rho_i\}$ with
efficiency
\begin{equation}
\begin{split}
P & = \sum_{i=1}^m \eta'_i\tr(\Pi_i\rho_i)\\
& = \sum_{i=1}^m
\eta'_i\sum_{k=1}^{n_i}\frac{\eta_{ik}}{\eta'_i}\tr(\Pi_i\sigma_i^k)\\
& = \sum_{i=1}^m\sum_{k=1}^{n_i}\eta_{ik}\tr(\Pi_i\sigma_i^k),
\end{split}
\end{equation}
which is also the efficiency of measurement operators $\{\Pi_i\}$
to unambiguously discriminate $\{S_i\}$. \hfill $\Box$

In some cases, a quantum system is secretly selected in some known
mixed states, which cannot be discriminated unambiguously. So it
is impossible to decide the exact state of the system
unambiguously. It is easy to conceive that if we only want to know
a certain range of states the system is in, we can perform a set
discrimination which may be able to be unambiguous. As shown in
Ref. \cite{Set}, this is true for pure states. However, for mixed
states, from Theorem \ref{Set}, we know that the problem of
unambiguously discriminating sets can be reduced to that of
unambiguously discriminating mixed states. So, the information we
can obtain from the system by using a set discrimination is the
same as that we obtain by discriminating the system among a fewer
number of some mixed states. Thus, unlike the problem of set
discrimination for pure states, which have many useful and unique
results, it is needless to deal with that of ''set discrimination
for mixed states'' specially.

\section{summary}
In this paper, we propose a method to transform each mixed state
being discriminated into the sum of two density matrices: one
matrix whose support space is linearly independent to each other
is called the ''core'' of the state, while the other one has no
contribution to discrimination. In this way, we can reduce the
problem of unambiguously discriminating mixed states into that of
unambiguously discriminating sets of pure states derived from the
''core''s. It is shown that a necessary and sufficient condition
for unambiguously discriminating mixed states is whether each of
their ''core''s is not zero. We also evaluate the efficiency of
unambiguous discrimination among mixed states, and present an
upper bound on it. Finally, we generalize the concept of set
discrimination to mixed states, and point out that the
unambiguous discrimination of mixed state sets is equivalent to
unambiguous discrimination of mixed states.


\begin{thebibliography}{9999}
\bibitem{dSum} In this paper, the sum of subspace means the space
in which any vector can be written as a linear combination of
vectors from those subspaces, such as
\begin{equation}
\sum_i S_i = \{\sum_i a_i\ket{v_i}\mid\ket{v_i}\in S_i\}.
\end{equation}
\end{thebibliography}
\end{document}